\documentstyle[floats,aps,epsf,prb]{revtex}
\begin{document}
\draft

\twocolumn[\hsize\textwidth\columnwidth\hsize\csname
@twocolumnfalse\endcsname

%
%
\title{Atomic structure of dislocation kinks in silicon}

\author{R.W. Nunes$^1$, J. Bennetto$^2$, and David Vanderbilt$^2$}
\address{$^1$Complex System Theory Branch, Naval Research Laboratory,
Washington DC, 20375-53459\\
and Computational Sciences and Informatics, George Mason University,
Fairfax, Virginia\\
$^2$Department of Physics and Astronomy, Rutgers University,
Piscataway, New Jersey 08855-0849\\}

\date{July 15, 1997}
\maketitle

\begin{abstract} 

We investigate the physics of the core reconstruction and associated
structural excitations (reconstruction defects and kinks) of
dislocations in silicon, using a linear-scaling density-matrix
technique. The two predominant dislocations (the 90$^{\circ}$ and
30$^{\circ}$ partials) are examined, focusing for the
90$^\circ$ case on the single-period core reconstruction.
In both cases, we observe
strongly reconstructed bonds at the dislocation cores, as suggested in
previous studies.  As a consequence, relatively low formation energies
and high migration barriers are generally associated with
reconstructed (dangling-bond-free) kinks.
Complexes formed of a kink plus a reconstruction defect are found to
be strongly bound in the 30$^\circ$ partial, while the opposite is
true in the case of 90$^\circ$ partial, where such complexes are found
to be only marginally stable at zero temperature
with very low dissociation
barriers.  For the 30$^\circ$ partial, our calculated formation
energies and migration barriers of kinks are seen to compare
favorably with experiment.  Our results for the kink energies on the
90$^\circ$ partial are consistent with a recently proposed alternative
double-period structure for the core of this dislocation. 

\end{abstract}

\pacs{61.72.Lk, 71.15.Pd, 71.15.Fv}

\vskip1pc]

\narrowtext

\section{Introduction}
\label{sec1}

Dislocations are of fundamental importance in the physics of
semiconductors, both from a mechanical and from an electronic point of
view. They are the carriers of plasticity in crystals and act as
trapping and scattering centers for electronic carriers.  A wealth of
experimental information is available about the properties of
dislocations in tetrahedrally bonded semiconductors.
\cite{hirsch,duesbery,alexan,gotts,farber,nikit,kolar} In Si,
the predominant slip systems are the 60$^{\circ}$ and the screw
dislocations oriented along $\left[110\right]$ directions in a
$\{111\}$ slip plane. Both are known to occur in the glide
configuration and to dissociate into pairs of partial dislocations
bounding a ribbon of intrinsic stacking
fault.\cite{hirsch,duesbery,alexan} Dissociation lowers the strain
energy and is made energetically favorable by the low energy of the
stacking fault in this material. (Evidence indicates that the above is
also true in the case of germanium and for III-V and II-VI
semiconductors.\cite{hirsch,duesbery}) The resulting 90$^{\circ}$ and
30$^{\circ}$ partials are believed to undergo core reconstruction,
which eliminates the unsaturated bonds, thus restoring the fourfold
coordination of the atoms at the cores. This picture is consistent
with the low density of dangling bonds, as suggested by EPR
measurements.\cite{hirsch,duesbery}

Dislocation motion occurs by nucleation and propagation of kinks along
the dislocation line. Due to thermal fluctuations or the action of an
applied stress, double kinks can be nucleated along the dislocation
line.  When these reach a critical separation, dissociation occurs and
the individual kinks propagate in opposite directions, thus generating
a displacement of a dislocation segment.\cite{h&l} A detailed
understanding of the atomic-scale structure of the kinks and the
barriers associated with their motion is thus of the greatest
importance.

In semiconductors, according to the theoretical model proposed by
Hirth and Lothe (HL),\cite{h&l} double-kink nucleation and the high
Peierls potential barrier to kink motion control the rate of
dislocation propagation. This is to be contrasted with the case of
metals, where kinks experience a very low barrier to motion, and the
rate is controlled by nucleation only. The HL model is often used in
the interpretation of dislocation mobility experiments, although the
occurrence of such high Peierls barriers has been questioned by some
authors.\cite{maeda} Furthermore, an alternative theoretical model has
been proposed in which dislocation motion is controlled by the pinning
of kinks by obstacles distributed along the dislocation
line.\cite{obst1,obst2} Recent experimental evidence suggests that the
barriers are indeed high, but experiments cannot clearly decide
between these two theoretical models.\cite{gotts,farber,nikit,kolar} A
complete microscopic picture is still lacking. Related issues, such as
the dependence of dislocation mobility on doping and the photoplastic
effect in semiconductors,\cite{hirsch} would also profit from a better
understanding of dislocation structure at the atomic level.

On the computational side, large-scale problems of this nature have
been mostly studied by using classical interatomic potentials.  Such
studies are not always reliable, since these potentials are often
unable to reproduce effects of intrinsic quantum-mechanical nature
such as bond reconstruction and Peierls or Jahn-Teller symmetry
breaking. For example, while the Stillinger-Weber~\cite{stwb} potential
has been used to study the core reconstruction and kinks of the
30$^{\circ}$ partial,\cite{bulatov} it fails to reproduce the
spontaneous symmetry-breaking core reconstruction of the 90$^{\circ}$
partial from the symmetric ``quasi-fivefold'' reconstruction.
\cite{bigger,nbv} A proper quantum-mechanical description of
the electronic structure is clearly needed. One is thus led to
consider tight-binding (TB) and {\it ab initio} methods.

Recent {\it ab initio} and TB theoretical works have concentrated on
such issues as the core reconstruction of the 90$^{\circ}$
partial,\cite{bigger,hansen} and the elastic interaction between
dislocations of a dipole in the shuffle~\cite{arias} and glide
sets.\cite{hansen} Using a relatively small supercell, one
first-principles study has determined a kink mobility barrier in the
30$^{\circ}$ partial,\cite{huang} but only one kink species was
studied, out of a very rich variety characteristic of this system. As
will become clear from the conclusions of the present work and from
Ref.~\onlinecite{bulatov}, the formation and migration energies of
other kinks are needed for a proper comparison with the experimental
results.  An important recent development is our prediction,\cite{bnv}
on the basis of classical-potential, tight-binding, and {\it ab
initio} calculations, that the reconstruction of the 90$^{\circ}$
partial in Si is not the above-mentioned symmetry-breaking structure,
as had generally been accepted in the theoretical literature.
\cite{bigger,nbv,hansen,chel84,markl83,markl94,jones80,jones93,%
heggie83,heggie93,oberg}
Instead, we proposed a
double-period (DP) reconstruction whose core structure is
reminiscent of that of a double-kink of minimal length that would form
on the core of the original single-period (SP)
structure.\cite{bnv} Cluster calculations on kinks and
solitons~\cite{foot-sol} in the SP reconstruction of the 90$^{\circ}$
partial have also been reported.\cite{jones80,heggie83,heggie93,oberg}
These calculations have identified many of the basic types of defects
in this system, but must be taken at a semi-quantitative level, since
they suffer from the lack of coupling of the defect local strain
fields with the lattice elastic fields.

To address properly the issues related to dislocation mobility, a
comprehensive study of dislocation kink structure and dynamics would
require the use of very large supercells, for which the application of
{\it ab initio} techniques is still computationally prohibitive.  In
view of this, the natural choice is the application of more efficient
quantum-mechanics based methods to study the electronic and structural
excitations in the dislocation cores.  In this work, we employ the
tight-binding-total-energy (TBTE) Hamiltonian of Kwon and
collaborators~\cite{kwon} to carry out a detailed atomistic study of
the atomic structure of both the 30$^{\circ}$ and the 90$^{\circ}$
partial dislocations in Si. To make these calculations tractable,
we use the linear-scaling or ``${\cal{O}}(N)$'' method of Li, Nunes
and Vanderbilt~\cite{lnv} to solve for the electronic-structure
contribution to energies and forces, enabling us to treat system sizes
up to $10^3$ atoms easily on a workstation platform.  Our work
addresses some of the fundamental issues associated with these two
systems. More specifically, we address the ground-state structural
properties of the dislocation cores and of defects in the core, such
as kinks and reconstruction defects (RD), as well as energy barriers
to motion of the various defects. 

In this work, when considering the 90$^{\circ}$ partial, we will
discuss only the SP reconstruction.  Despite the fact that this is not
the correct ground state for this dislocation in Si, we hope that
understanding this somewhat simpler system will help us in the study
of the myriad of defects in the more complicated ground-state DP
reconstruction, to which the former is related.\cite{bnv} Moreover, we
should keep in mind that the 90$^\circ$ partial is equally important
in other materials, such as germanium (Ge), diamond (C), and the III-V
and II-VI semiconductors.\cite{hirsch,duesbery,alexan} Preliminary
calculations,\cite{bnv} using a Keating model,\cite{keating} indicate
that in C the SP reconstruction is more likely to be lower in energy,
while Ge, like Si, would prefer the DP structure. More accurate
calculations are needed to reveal which of the two reconstructions
would be favored in each case. Therefore, the study of the SP
structure is still important from a theoretical point of view.

The paper is organized as follows. The next section contains the
technical details of the calculations we performed. In
Sec.~\ref{sec3}, we discuss our results for the core reconstruction
and related defects in the 30$^\circ$ partial dislocation. Our main
results for the SP reconstruction of the 90$^\circ$ partial are
described in Sec.~\ref{sec4}. Finally, in Sec.~\ref{sec5}
we summarize the main conclusions and results, and compare our kink
energies and barriers with the available experimental results. In
particular, we will argue that our results appear to be consistent with
the HL theory of dislocation glide.

\section{Technical details of the calculations}
\label{sec2}

We use the TBTE parameters of Kwon {\it et al.},\cite{kwon} which
describe well the acoustic phonon modes and elastic constants of Si,
thus being adequate to describe the strain fields associated with the
dislocation cores and related defects. Owing to its good
transferability between different crystal structures, ranging from
diamond to FCC, this Hamiltonian is also expected to give a good
description of the coordination defects in the present study.  The
${\cal{O}}(N)$ method of Li {\it et al.}~\cite{lnv} is used to solve
for the electronic structure contribution to the energies and
forces. For the density matrix, we initially work at a real-space
cutoff $R_c=$ 6.2 \AA\ on the range of the density matrix used in the
tests presented in Ref.~\onlinecite{lnv} for the ${\cal{O}}(N)$
method. In a second stage, we improve the convergence of our results
by further relaxing the ionic positions and the electronic structure
with a larger cutoff value of $R_c =$ 7.3 \AA.  The numerical
minimization of the ${\cal{O}}(N)$ functional was carried out by the
conjugate-gradient algorithm, with the internal line minimization
performed exactly. To obtain the right number of electrons, the
chemical potential is adjusted iteratively, in each case. Usually,
this procedure has to be repeated only at the initial steps of the
structural relaxation procedure, after which the chemical potential
converges to the adequate value and remains constant.  Ground-state
structures were computed by allowing all atomic coordinates to relax
fully (average forces less than 0.1 meV/\AA).

Our supercells are chosen with the dislocation direction
(corresponding to a $\left[1{\overline 1}0\right]$ crystalline
direction) lying on the $y$-axis. The glide plane (which contains a
stacking fault) is normal to the $\left[111\right]$ direction and
coincides with the $xy$ plane of the cell.  (Fig.~\ref{core-30} shows
the glide plane of the 30$^\circ$ partial dislocation, with the
crystalline directions indicated.)  The $z$ direction of the cell is
thus parallel to the $\left[111\right]$ direction.  Each supercell
contains two dislocations having opposite Burgers vectors (a
dislocation dipole), which allows us to use periodic boundary
conditions. Supercell vectors are chosen such as to array the
dislocations in a quadrupole configuration, as suggested in
Ref.~\onlinecite{bigger}, to avoid the spurious shear strains
associated with the minimal dipole cell.

To ensure the convergence of our calculations with respect to
supercell size, we used three different cells, containing 216, 264,
and 308 atoms respectively, for the simulation of the reconstructed
core of the 30$^\circ$ partial dislocation. Each cell corresponds to a
slab of atoms at a 60$^\circ$ angle with respect to the dislocation
direction, including twice the lattice period in that direction, to
allow for the period-doubling reconstruction of the 30$^\circ$
partial.  The two parameters characterizing the geometry of each cell
are the separation between the two dislocations in the glide plane
(i.e., the width of the stacking fault) within a given unit cell, and
the distance between the periodic-image dipoles along the $z$
direction. In our calculations, these distances are, respectively,
15.0~\AA\ and 18.8~\AA\ for the 216-atom cell, 18.3~\AA\ and 18.8~\AA\
for the 264-atom cell, and 18.3~\AA\ and 21.9~\AA\ for the 308-atom
cell.

The supercells for the computation of defect energies are obtained by
repeating the core slabs several times along the dislocation
direction.  The defect energies we quote are referred to the
corresponding supercell containing defect-free (but reconstructed and
fully relaxed) dislocations. For the kinks in the 30$^\circ$ partial,
each of the core slabs were repeated three times (two and a half 
times for the RD, and three and a half times for the kink-RD 
complexes) along the dislocation 
direction, so that the defect-defect separation along the line was 
19.2~\AA\ or larger, depending on the type of defect. 
Below, we describe the procedure we used for the computation of defect
energy barriers. Because of the higher computational demands involved,
in this case we employed only the smaller cells (three times the
216-atom slab for kinks and two and a half times the same slab for the
RD).

Table I in the next section illustrates the convergence of our results
with respect to dislocation separation. As a further check, we also
computed the energies of the core and of one the kinks (the LK kink,
as described below) with an even larger slab, consisting of 600
atoms for the reconstructed core (1800 atoms for the defect). In this
case, dislocation distances are 24.9~\AA\ in the $xy$ plane and
31.4~\AA\ in the $z$ direction. The change in defect energy with
respect to the 308-atom slab was only $\sim$0.02~eV.  To test the
effect of defect interaction, this kink was studied with a larger
kink-kink separation (with the smallest slab repeated eight times),
which produced a change of only $\sim$0.01~eV in the
energy. Therefore, we consider our calculations to be converged within
0.03~eV with respect to core-core and defect-defect interactions.
To estimate the error involved our choice of cutoff for the density
matrix, kink and core energies were computed using a larger cutoff
($R_c=$ 8.1 \AA). The kink formation energy changed by only 
$\sim$0.06~eV, which justifies our choice of cutoff. 
From these results we can also estimate that our defect energy
barriers are converged within 0.3 eV.
The supercells used for the study
of the 90$^\circ$ partial, are as described in
Refs.~\onlinecite{nbv,bnv}. In this case, despite the fact that we are
only interested in the qualitative nature of our results, our values
are well converged, with dislocation distances on the order of
26.6~\AA, and defect-defect separations of at least $\sim$13.4~\AA.
(As in the case of the 30$^\circ$ partial, barriers are computed using
smaller cell sizes, corresponding to a dislocation separation of
13.3~\AA.)

Barrier energies were calculated by identifying the 3$N$-dimensional
configuration-space vector ${\bf R}_{12}={\bf R}_2-{\bf R}_1$ pointing
from one equilibrium position ${\bf R}_1$ of the defect to a
neighboring position ${\bf R}_2$, and defining a reaction coordinate
$Q = {\bf R} \cdot {\bf R}_{12}$ measuring the progress from ${\bf
R}_1$ to ${\bf R}_2$.  Then, for a series of values of this
coordinate, we computed the energy with this coordinate fixed and all
others fully relaxed. This approach is efficient in simple cases, but
we find that it often fails to converge to the saddle-point
configuration when the reaction path~\cite{foot-path} makes sharp
angles with respect to ${\bf R}_{12}$.  In these cases, we can usually
find two configurations, ${\bf R'_1}$ and ${\bf R'_2}$, near the
saddle-point, with nearly the same value of $Q$ but with opposite
forces along ${\bf R}_{12}$.  By exploring the space spanned by ${\bf
R}_{12}$ and $\left({\bf R'_2} - {\bf R'_1}\right)$
while allowing all other coordinates to relax, we were able
to determine the energy barriers with good accuracy for all
cases studied (average forces less than 1.0 meV/\AA).

\section{The 30$^\circ$ partial dislocation} 
\label{sec3}

\subsection{Core reconstruction}
\label{sbsec3.1}

\begin{figure}
\epsfxsize=2.8 truein
\centerline{\epsfbox{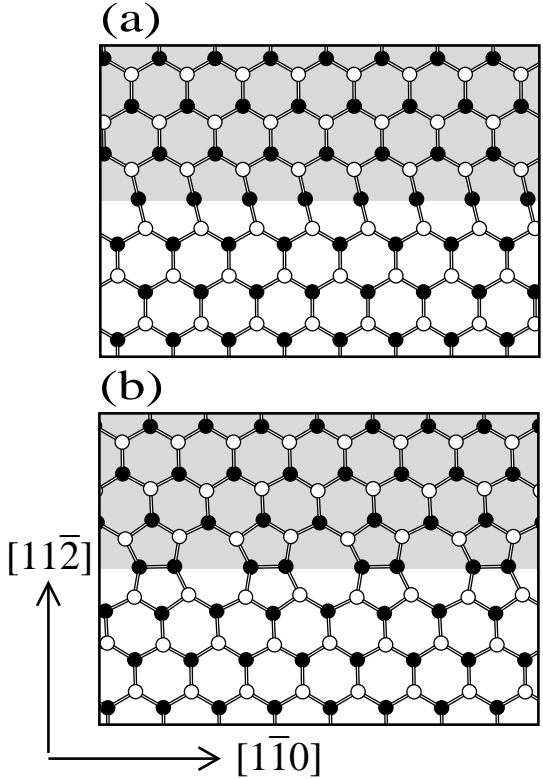}\quad}
\vskip 0.20truein
\caption{(a) Unreconstructed core of the 30$^\circ$ partial
dislocation, viewed from above the $(111)$ slip plane.  Shaded region
indicates stacking fault.  Black (white) atoms lie below (above) the
slip plane. (b) Same view of the double-period reconstructed
structure. Crystalline directions are also shown.}
\label{core-30}
\end{figure}

In Fig.~\ref{core-30}(a), a top view of the atomic structure of the
unreconstructed 30$^{\circ}$ partial in the glide plane is shown.  The
shaded area represents the stacking fault, and the dislocation line is
indicated by the boundary between shaded and unshaded areas. The
crystalline directions are also displayed.  Atoms shown as white
(black) are above (below) the glide plane; each atom is bonded to
another either above or below it, and these are not shown in the
picture. Thus, fourfold coordinate atoms have three of their bonds in
the plane of the figure. The atoms at the core of the defect are
threefold coordinated, with a dangling bond lying nearly parallel to
the dislocation line.  In Fig.~\ref{core-30}(b) we show a
reconstruction in which the fourfold coordination of the atoms at the
core is restored by atoms bonding in pairs along the line, leading to
a doubling of the period in that direction.  This reconstruction is
well accepted as being the ground-state of the 30$^{\circ}$ partial,
and has been discussed theoretically by other
authors.\cite{markl83,northrup,chel82,bulatov} In
Ref~\onlinecite{bulatov}, it was found to be 0.21 eV/\AA\ lower in
energy than the unreconstructed structure, using a Stillinger-Weber
potential.  We find a higher value of 0.36 eV/\AA\ for the
reconstruction energy.

\begin{table}
\caption{Formation energy of defects in the 30$^\circ$ partial
dislocation, in eV. Defect energies are referred to a defect-free
dislocation core. TB results for three supercell sizes are shown. For
the PSD, supercells contain 5/6 of the number of atoms shown. For the
LK, in the third column we indicate in parenthesis the formation energy
computed with a 1800-atom cell. Fourth column contains Keating energy
computed for the largest cell (with ``Keating + 0.4 eV'' numbers in
parenthesis). }
\begin{tabular}{lcccc}
 &648 atoms   &792 atoms
 &924 atoms   &Keating \\
\hline
PSD        &1.35    &1.32     &1.33    \\
LK         &0.52    &0.37     &0.35 (0.33)   &-0.06 (0.34) \\
LK$'$        &0.97    &0.81     &0.76    &$\;$0.44  (0.84) \\
RK         &0.93    &1.20     &1.24    &$\;$1.00  (1.40) \\
RK$'$        &1.64    &1.84     &1.85    &$\;$1.30  (1.70) \\
\end{tabular}
\end{table}

A look at the distribution of bond lengths for this structure shows
that the reconstruction is indeed strong, with maximum bond-length
deviations of only 3.0\% (maximum and minimum bond lengths are 2.423
\AA\ and 2.308 \AA, respectively) with respect to Si bulk values
(2.351 \AA). The core energy is mostly due to the strain associated
with bond-angle distortions at the core of the defect, with
bond angles ranging between $\sim$90$^\circ$ and $\sim$126$^\circ$
($109.5^\circ$ is the bulk value). No mid-gap levels are expected for
this structure, in accordance with the EPR
evidence.\cite{hirsch,duesbery,alexan} 

A rich variety of core defects is associated with this reconstruction,
including kinks and RDs, and complexes of these basic types.  A very
extensive study of these defects is found in Ref.~\onlinecite{bulatov},
including structural features and energetics under
a Stillinger-Weber potential. To a large extent, our study of this
specific dislocation relies on this previous study, adding to it the
benefits of a quantum-mechanical treatment of the electronic
structure. More specifically, the defects considered in this work
are the ones identified in Ref.~\onlinecite{bulatov}.
As we proceed, it will be seen that some of our results differ
qualitatively from those in Ref.~\onlinecite{bulatov}, and also that we
find a better agreement with the experimental results.

\subsection{Phase switching defect (PSD)}
\label{sbsec3.2}

\begin{figure}
\epsfxsize=2.8 truein
\centerline{\epsfbox{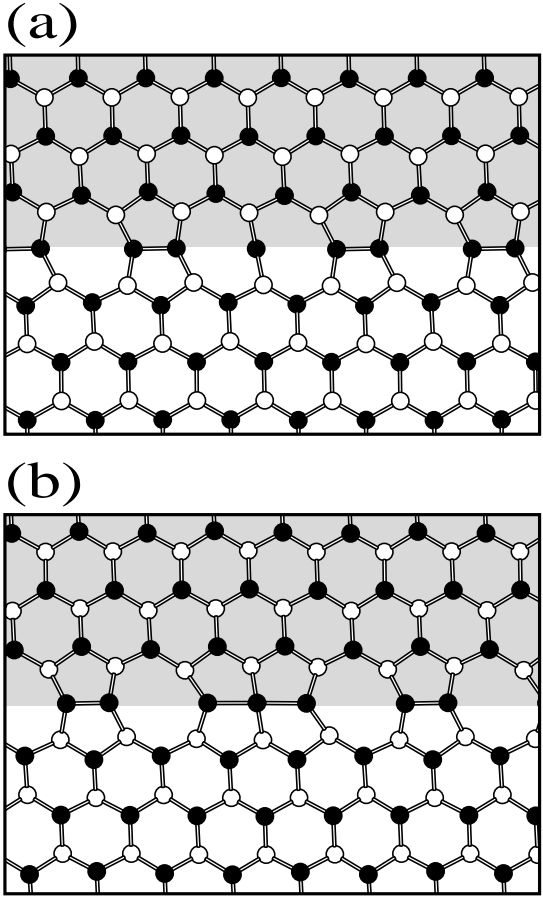}}
\vskip 0.20truein
\caption{(a) Core structure of a phase-switching defect (PSD), which
is a reconstruction defect in the core 30-partial dislocation. The
phase of the reconstructed bond along the dislocation line is
switched, going through the defect. (b) Saddle-point configuration for
the propagation of a PSD along the core.}
\label{PSD}
\end{figure}

Fig.~\ref{PSD}(a) shows a RD associated with the core of the
30$^\circ$ partial. We shall refer to this defect as a phase switching
defect (PSD).\cite{foot-sol} The existence of such defects has been
hinted at since the realization that the core of the partials might
undergo reconstruction.\cite{hirsch,jones80} We computed the energy of a
fully relaxed PSD by repeating the atom slabs five times along the
$\langle 110\rangle$ direction, and introducing one PSD in each
dislocation line. Our value for the PSD formation energy is 1.32 eV
(see Table I below), which is somewhat higher than the value of 0.81
eV obtained in Ref.~\onlinecite{bulatov}. We believe our result to be
more reliable, given the quantum-mechanical nature of our approach, in
particular for a defect containing a dangling bond. To a first
approximation this defect can be understood as a $p$ dangling-bond
defect, which indicates that a formation energy on the order of 1 eV
(roughly the bond-breaking energy in bulk Si) is to be expected. The
exact value is determined by the relaxation of the atoms surrounding
the defect.

We also computed the migration barrier for the propagation of the PSD
along the dislocation direction.  The relaxed structure of the barrier
(saddle-point) configuration is shown in Fig.~\ref{PSD}(b).
In this case, the symmetry between adjacent positions of the defect
along the line indicates that the saddle-point configuration is at
the halfway position. It was somewhat surprising to find that even
in this case, we had to resort to a two-dimensional reaction
coordinate as described Sec.~\ref{sec2}. Our
saddle-point configuration, with an energy barrier of 0.3 eV, is very
similar to that in Ref.~\onlinecite{bulatov}. In can be seen that the
atom at the center becomes fivefold coordinated, which leads to a
smooth process of bond substitution as the PSD propagates to the right.
This explains the low energy barrier involved in this process.

\subsection{Kinks}
\label{sbsec3.3}

\begin{figure}
\epsfxsize=2.8 truein
\centerline{\epsfbox{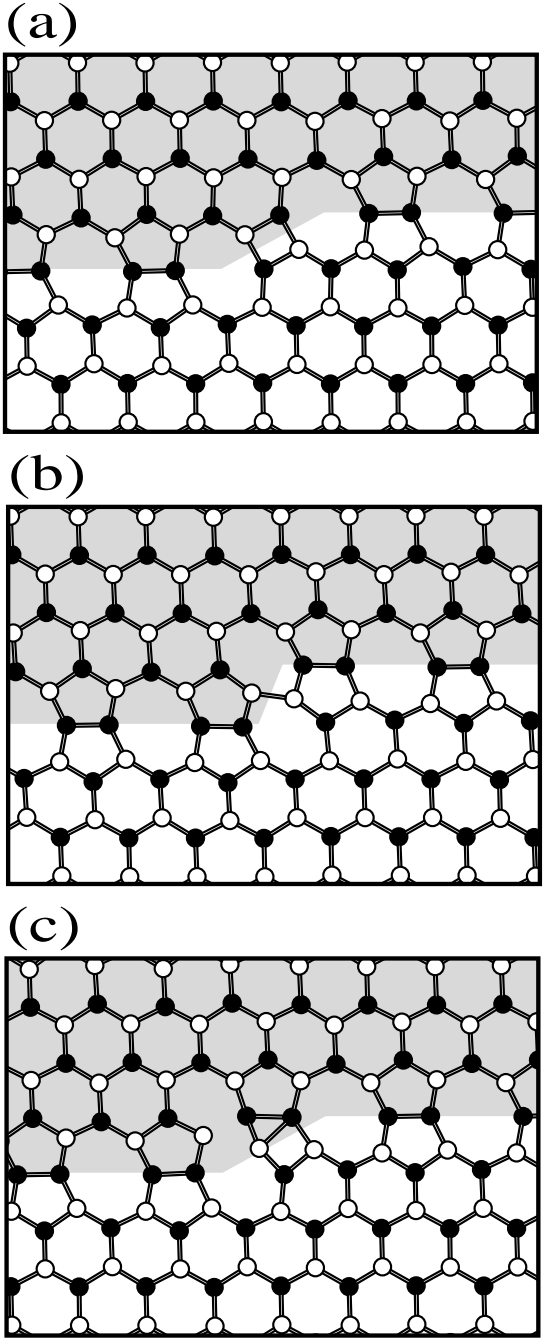}}
\vskip 0.20truein
\caption{Core structure of the left kinks in the 30$^\circ$ partial,
and associated transition state. Kink notation is explained in the
text. (a) LK kink. (b) LK$'$ kink. (c) Transition state for the LK
$\rightarrow$ LK$'$ transformation.}
\label{LK}
\end{figure}

\begin{figure}
\epsfxsize=2.8 truein
\centerline{\epsfbox{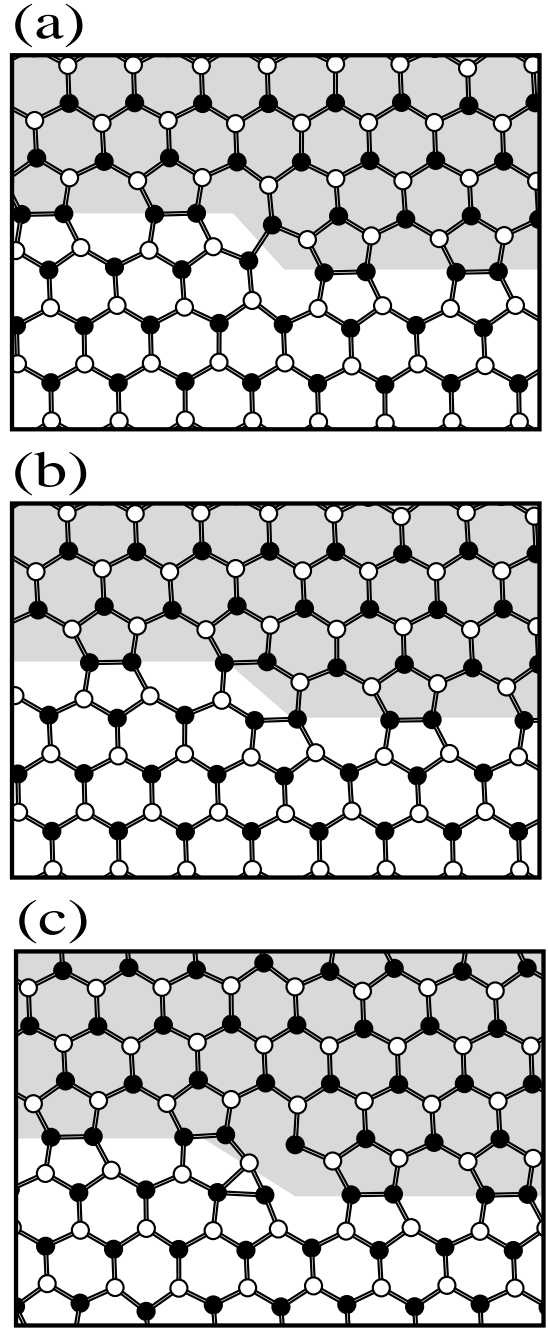}}
\vskip 0.20truein
\caption{Core structure and transition state of right kinks in the
30$^\circ$
partial.  Kink notation is explained in the text. (a) RK kink. (b)
RK$'$ kink. (c) Transition state for the RK $\rightarrow$ RK$'$
transformation.}
\label{RK}
\end{figure}

The period doubling of the reconstructed core gives rise to a
multiplicity of kinks in this system. Two distinct families of such
defects appear, depending on whether the dislocation ``kinks'' to the
left (Fig.~\ref{LK}) or to the right (Fig.~\ref{RK}). The period
doubling of the core introduces a choice of phase of the
core reconstruction both ahead of, and behind, the kink.
Of the four configurations generated in this way, two of them
(those necessarily containing a coordination defect) will be
classified as PSD-kink complexes, and will be considered in
Sec.~\ref{sbsec3.4}.  The remaining two configurations will
be classified as ``pure'' kinks and are considered here.  The
two left kinks LK and LK$'$ are shown in Figs.~\ref{LK}(a-b),
while the two right kinks RK and RK$'$ are shown in Fig.~\ref{RK}(a-b).

The energies for each type of kink were computed using the TBTE
Hamiltonian, as well as with a classical Keating model,\cite{keating}
with the parameters proposed in Ref.~\onlinecite{qianchadi}, in order
to look at the local-strain contributions to the energy of each
defect. In Table I, we show the TBTE results for each of the three
slabs described in section~\ref{sec2}, along with the Keating results
for the 924-atom cell. For one of the kinks, the energy computed with
a 1800-atom slab is also shown in parenthesis; note the convergence of
these results with respect to cell size.
Next, we discuss the results for each kink family in more detail.

\subsubsection{Left kinks}

The left kinks LK and LK$'$ are shown in Fig.~\ref{LK}, together with
the saddle-point configuration for the LK $\rightarrow$ LK$'$
translational
motion. The energies, as given in Table I, show that reconstruction
produces low energy kinks in this case, as compared to the energy of
the unreconstructed PSD defect. At first sight, the formation energy
for these reconstructed defects is expected to be mostly associated
with the local strain at the kink cores. The Keating-model results can
give us an estimate of these local-strain effects. The LK is found to
add no additional strain on that imposed by the core reconstruction
itself, as can be seen by its slightly negative energy.  On the other
hand, the LK$'$ kink is found to have a Keating formation energy of 0.44
eV.  It is interesting to note that the {\it relative} Keating energy of
the two left kinks is in good agreement with the TB results.  This is
actually true for all four kink types, as can be seen be comparing
differences
of the Keating energies in the fourth column of Table I. In the last
column, we add a constant shift of 0.4 eV to each Keating energy. It
is evident that the ``Keating + 0.4 eV'' results are reasonably close to
the TB ones. In view of this, we conclude that a roughly constant
band-structure energy of 0.4 eV is associated with each kink.

For the saddle-point configuration in Fig.~\ref{LK}(c) we computed
an energy barrier of 1.52 eV. This result is in very good agreement
with experimental estimates. In our concluding section, we will
discuss more extensively the significance of our results in light of
the available experimental evidence. Here, we note that such a high
barrier can be understood by the presence of severe bond-bending and
stretching distortions at the core of the defect, along with the
presence of malcoordinated atoms.  Bond angles as small as
50.4$^\circ$ are found, as well as bonds stretching to 2.80 \AA.

\subsubsection{Right kinks}

Shown in Fig.~\ref{RK} are the two kinks of the right family, RK and
RK$'$, together with the saddle-point corresponding to the RK
$\rightarrow$ RK$'$ reaction.  Despite the fact that both kinks are
fully reconstructed, the formation energies of 1.24 eV for RK and 1.85
eV for RK$'$ are surprisingly high. Again, note the agreement between
the Keating values and the TB ones, after adding a constant shift of
0.4 eV to the former. No single structural feature of the right kinks
could be traced in order to explain the unexpected formation energies.
The minimum and maximum distortions of bond lengths and angles do not
vary drastically among the four kink types.

\begin{figure}
\epsfxsize=3.3 truein
\centerline{\epsfbox{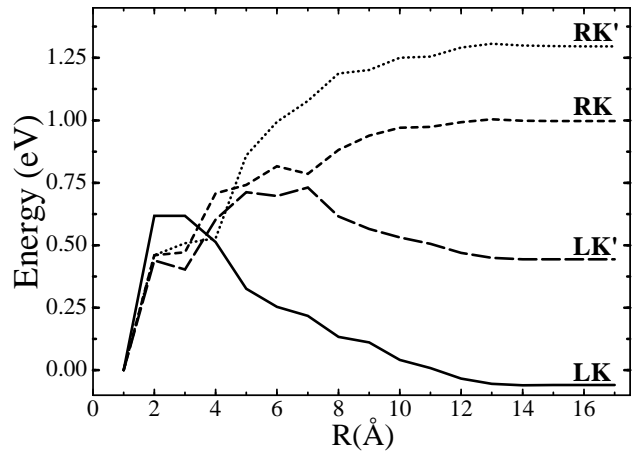}}
\vskip 0.20truein
\caption{Keating energy for 30$^\circ$-partial kinks. Energy $E(R)$
is the sum over all atoms within a distance $R$ from the
dislocation core. Corresponding core energy is subtracted to yield
defect energies.}
\label{keating}
\end{figure}

To help us better understand these results, we observe that the
Keating energies can be decomposed in an atom-by-atom basis. Bond
bending energies, associated with changes in the angle between two
bonds, are assigned to the vertex atom, and half of the bond stretching
energy of a given bond is assigned to each of the two participating
atoms.  To examine the nature of the strain fields associated with
each kink type, using our largest cells (924 atoms), we looked at
these atomic energies integrated over shells of atoms defined by their
distance from the core of the defect. (Since our supercells contain
two cores and thus two defects, we always choose the shortest distance
to a defect.)  The integrated energies are then defined by
\begin{equation}
E^d(R)= \sum_{R_i \leq R} E^d_i(R_i)-E^c_i(R_i)\;,
\end{equation}
where the Keating energy $E^c_i(R_i)$ of each atom in a correspondent
kink-free supercell (containing only the dislocation dipole) is
subtracted, and we sum over all atoms within a distance $R$ from the
kink.  The results are shown in Fig.~\ref{keating}. We see that the
kink energies are determined by the medium-range behavior of the
associated strains. At short range ($R<3.0$\AA) the LK kink is
actually the highest in energy. As we advance away from the core of
the kinks, the energies only approach their final relative values at a
distance of about $R=$10.0\AA.

As in the case of the left kinks, a look at the saddle-point
configuration shows that the rather high migration barrier of 2.03 eV
for the right kinks is associated with the drastic bond distortions
and malcoordination of atoms at the core. Note that this barrier is
substantially higher than the 1.52 value we obtained for the left
kinks, leading to a physical picture of ``fast'' and ``slow''
plasticity carriers for the 30$^\circ$ partial dislocation. In our
concluding section, we discuss this point further.

Here, it is worth pausing to compare our results with those in
Ref.~\onlinecite{bulatov}. Individual kink energies are not obtained
in their work, since in all their calculations the supercells
contained a double kink (two kinks, one of each family). Therefore, we
cannot compare our kink energies directly with their results. In their
procedure, what is computed are the relative energies of kinks within
each kink family, assuming the LK and RK to have the same energy.
A first aspect to be pointed out is that the above assumption of
degeneracy between the LK and RK kinks is in sharp disagreement with
our findings. In agreement with our work, they find an energy
difference of $\sim$0.4~eV between the two left kinks. On the other
hand, while our results indicate that the two right kinks also differ by
$\sim$0.4~eV, they find these two kinks to be almost degenerate, with
energies differing by 0.07~eV only. 
Our kink migration barriers are substantially higher, despite
the fact that the associated saddle-point configurations seem to be
very similar with those identified in their work. Below, our results
will be seen to compare more favorably with experimental estimates of
the kink barriers.

\subsection{PSD-kink complexes}
\label{sbsec3.4}

Kinks and PSDs can be considered as the fundamental types of
excitations in the dislocation cores. Important
structural features and modes of dislocation dynamics can also be
associated with the complexes formed by these basic defect
types. Moreover, since RDs such as the PSD are
malcoordinated (thus acting as weak links in the reconstructed core),
they may act as preferential sites for the nucleation of double kinks,
as suggested by Heggie and Jones.\cite{hirsch,heggie83} Possibly, a
PSD-kink complex could result from such a nucleation process, as the
double kink expands and eventually dissociates into single kinks.
Therefore, it is important to understand the structure and energetics
of these complexes. 

\begin{table}
\caption{Formation energy of defect complexes in the 30$^\circ$
partial dislocation, in eV. Two different states are considered for
each complex (notation is explained in the text). Binding
energy for the largest cell is indicated in the last column.}
\begin{tabular}{lcccc}
 &756 atoms   &924 atoms
 &1078 atoms   &Binding energy \\
\hline
LC(0)        &1.11    &0.97     &0.88    &0.80 \\
LC(1)        &1.78    &1.66     &1.58    &$\;$-\\
RC(0)        &1.90    &2.09     &2.15    &0.42 \\
RC(1)        &2.43    &2.55     &2.64    &$\;$- \\
\end{tabular}
\end{table}

\begin{figure}
\epsfxsize=3.3 truein
\centerline{\epsfbox{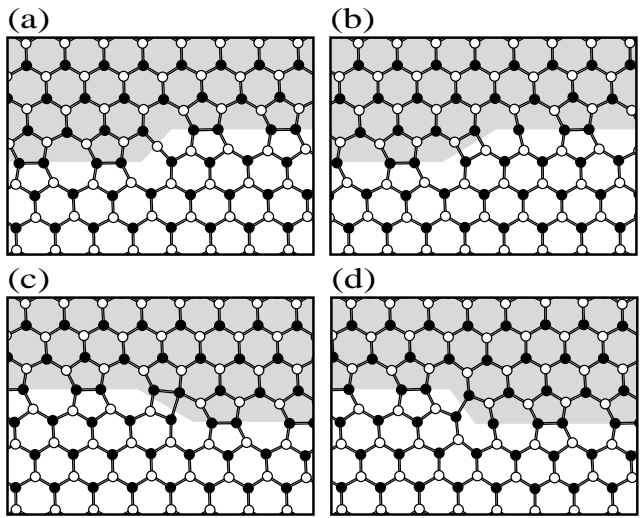}}
\vskip 0.20truein
\caption{Core structure of kink-PSD complexes in the 30$^\circ$
partial. In each case, two states of the complex are considered, as
explained in the text.  (a) LC(0) = LK + PSD at zero separation. (b)
LC(1) = LK + PSD one lattice period apart. (c) RC(0) = RK + PSD at
zero separation. (d) RC(1) = RK + PSD one lattice period apart.}
\label{cmplx}
\end{figure}

Here, we consider the energetics of the PSD-kink complexes. The
important questions concern whether or not these complexes form bound
states, as well as the associated binding energies and migration
barriers.  We considered each of the PSD-kink complexes in two
configurations, as shown in Fig.~\ref{cmplx}.  The left complex (LC =
LK + PSD) is shown in the state of closest approach, LC(0),
Fig.~\ref{cmplx}(a), in which the two constituents overlap and cannot
be distinguished; and in an extended state, LC(1),
Fig.~\ref{cmplx}(b), in which the PSD and the kink have been separated
to adjacent positions.  The corresponding right-complex cases RC(0)
and RC(1) are shown in Fig.~\ref{cmplx}(c) and (d), respectively. In
Table II we show our results for the energies of these four
configurations, where it can be seen that the PSD binds strongly with
both the left and the right kinks, in agreement with
Ref.~\onlinecite{bulatov}. Contrary to what is found in
Ref.~\onlinecite{bulatov}, our results indicate the LC to be more
strongly bound than the RC.
From the binding energies and the energies of these more
extended configurations, we obtain a lower bound of 0.80 eV (LC) and
0.49 (RC) for the dissociation barrier of these bound states.
Below,
these results will be shown to be in sharp contrast with those for
kink-RD complexes in the SP reconstruction of the 90$^\circ$ partial
dislocation, which are found to be unstable.  Finally, we note that
the energy of the LC is lower than that of the PSD, making the former
the more likely site for unpaired electrons in the core of the
30$^\circ$ partial.

\section{The 90$^\circ$ partial dislocation} 
\label{sec4} 

\subsection{Core reconstruction}
\label{sbsec4.1} 

\begin{figure}
\epsfxsize=3.3 truein
\centerline{\epsfbox{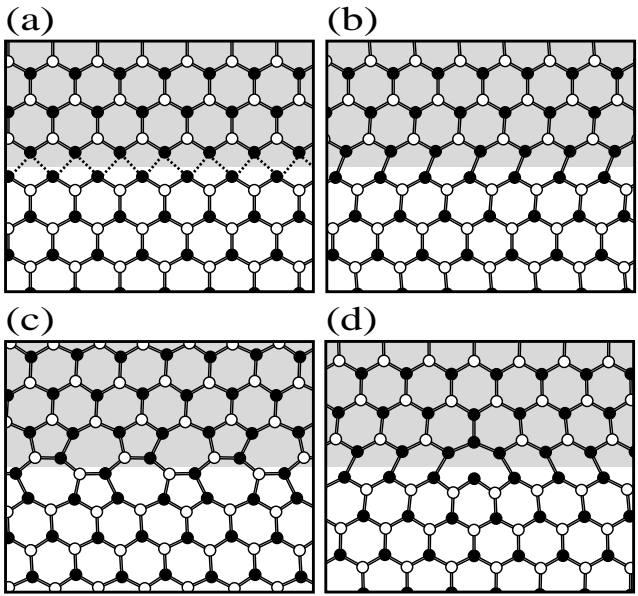}}
\vskip 0.20truein
\caption{Models for core reconstruction of the 90$^\circ$ partial
dislocation.  (a) Symmetric QF reconstruction. (b) Symmetry-breaking
SP structure.  (c) Ground state symmetry-breaking DP structure. (d)
Reconstruction defect or DSD in the SP core.}
\label{core-90}
\end{figure}

Considerable theoretical effort has been devoted to the study of the
90$^\circ$ partial dislocation.
\cite{bigger,nbv,hansen,chel84,markl83,markl94,jones80,jones93,%
heggie83,heggie93,oberg}
Basically, two types of core reconstruction have been
considered. These are the symmetric quasi-fivefold (QF) and the
symmetry-breaking SP reconstructions shown in Fig.~\ref{core-90}(a)
and (b), respectively. Both preserve the original periodicity of the
lattice along the dislocation direction. The latter structure has been
found to be lower in energy.  It was thus commonly assumed to be the
ground state in Si and other semiconductors, and the bulk of studies
of core excitations has relied upon this assumption. Recently, we
proposed an alternative solution for the ground-state in Si,\cite{bnv}
where a period-doubling symmetry-breaking structure, seen in
Fig.~\ref{core-90}(c), is shown to have lower energy than the SP
one. As a consequence, the study of core excitations and the related
issue of dislocation mobility have to be re-addressed. We are
currently undertaking this task, and the results will be published
elsewhere.

Nevertheless, we note that the DP structure is closely related to the
SP one, being obtained by inserting alternating kinks in the core of
the latter.\cite{bnv} Therefore, understanding the defect structure of
the simpler SP core may prove useful to the study of the rather large
number of core defects of the DP reconstruction. In this section, we
summarize our main results for the core and related defects of the SP
structure.

The SP core has two degenerate ground states, depending on the
direction of the symmetry-breaking bonds. By convention, we denote the
configuration in Fig.~\ref{core-90}(b) as the ``right'' reconstruction,
from which we can obtain the ``left'' state by applying the broken
mirror operations [the ones that are unbroken in the QF core in
Fig.~\ref{core-90}(a)]. We find the SP core to be 0.18 eV/\AA\ lower
in energy than the QF one. This result is in good agreement with
previous TB~\cite{hansen} (0.18 eV) and LDA~\cite{bigger} (0.23 eV)
works.  In Ref.~\onlinecite{bigger}, it was found that symmetry
breaking occurs spontaneously, a result that is confirmed by our model.
In our calculations, the reconstructed bonds are stretched 3.0\% with
respect to the perfect crystal values (2.5\% in
Ref.~\onlinecite{bigger}), and the minimum and maximum bond angles are
97$^{\circ}$ and 135$^{\circ}$, respectively (96$^{\circ}$ and
138$^{\circ}$ in Ref.~\onlinecite{bigger}).  Core defects are
considered next.

\label{sbsec4.2}

\subsection{Direction switching defect (DSD)}

Symmetry breaking in the SP core gives rise to a RD
in which the direction of the bonds is switched, as
shown in Fig.~\ref{core-90}(d). We shall refer to this defect as a
direction switching defect (DSD).\cite{foot-sol}  Note that, like the
30$^\circ$-partial PSD, this defect contains a dangling bond, which
explains its formation energy of 1.45 eV. Our result is in reasonable
agreement with the 1.2 eV value obtained in the cluster calculations
of Ref.~\onlinecite{heggie93}. For the DSD motion, we computed an energy
barrier of only 0.04 eV for the propagation between two adjacent
equilibrium positions.  Given such a small barrier, the DSD is
expected to be extremely mobile even at low temperatures. As a test,
we performed a molecular dynamics simulation on a supercell having a
pair of DSD defects, initially separated by 9.6 \AA, on
an otherwise defect-free partial dislocation. Remarkably, at a
temperature of only 50~K, recombination of the pair took
place after only 1.3 ps. Unlike PSDs in the 30$^\circ$ partial, such
highly mobile DSDs do not bind strongly with kinks to form DSD-kink
complexes, as explained below.

\subsection{Kinks}

\begin{figure}
\epsfxsize=2.8 truein
\centerline{\epsfbox{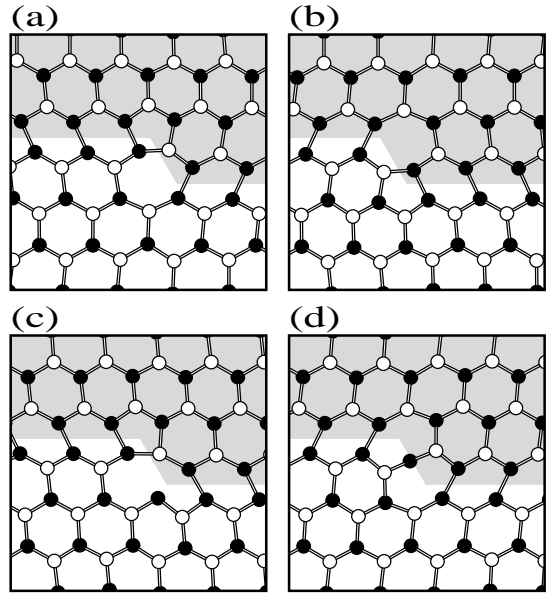}}
\vskip 0.20truein
\caption{Core structure of kinks and DSD-kink complexes in the SP core.
See text for notation. (a) LR kink. (b) RL kink. (c) LL complex = LR +
DSD.  (d) RR complex = LR(RL) + DSD.}
\label{defect-90}
\end{figure}

It would be possible to define left (LK) and right (RK) kinks in
the case of the 90$^\circ$ partial, just as for the 30$^\circ$ partial.
However, in the 90$^\circ$ case, each LK is directly related to
a corresponding RK by application of a mirror symmetry.  (This was
not true for the 30$^\circ$ partial, where the mirror symmetry
was absent from the outset.)  Thus, for the 90$^\circ$ partial,
we shall restrict the discussion to right kinks only.  Moreover,
we will now use the notation `L' and `R' in a completely
different way, namely, to denote the direction of the core
reconstruction on either side of the kink.  Referring to
Fig.~\ref{defect-90}(a), the reconstruction will be said to
tilt to the `left' and to the `right' on the left and right
sides of the kink, respectively.  Hence, we call this a left-right
(LR) kink, the notation following accordingly for the other
defects.

We compute the sum of the energies of the LR and RL kinks shown in
Fig.~\ref{defect-90}(a) and (b), to be 0.24 eV only.  The RL and LR
kinks are structurally quite similar; they would be related by a
two-fold rotation axis normal to the plane of Fig.~\ref{defect-90}, if
it were not for the fact that a stacking fault exists on one side but
not the other.  Thus we expect the energies of the two kinks to be
similar, and assign the average energy of 0.12 eV to each. The rather
low formation energy can be seen as another indication of the DP core
structure, since even individual kinks add little strain over that
imposed by the SP core itself. In the formation of the DP core, this
additional strain is more than compensated for by the attraction
between the LR and RL kinks. In the present context, this result only
shows that our previous results for kink energies in
Ref.~\onlinecite{nbv} were severely under-converged with respect to the
dislocation interaction in the cell. We also computed an energy
barrier of 1.62 eV for the motion of the LR and RL kinks. As is the
case for reconstructed kinks in the 30$^\circ$ partial, such large
energy barriers are associated with the existence of malcoordinated
atoms and severe bond distortions at the core of the kink.

\subsection{DSD-kink complexes}

There are two additional kink-type defects associated with the SP
reconstruction of the core. These are the RR and the LL defects, shown
in Figs.~\ref{defect-90}(c) and (d).  We prefer to regard these as
complexes of a LR or a RL kink together with a DSD.  Two LL complexes
are possible (only
one is shown in Fig.~\ref{defect-90}), and they share the same
``quasi-symmetry'' that the LR and RL kinks do, differing only by the
position of the fivefold and dangling-bond-containing rings with
respect to the stacking fault.  In contrast with complexes in the
30$^\circ$ partial, these complexes appear to be either unstable or
marginally stable against the emission of a DSD, as discussed in
Ref.~\onlinecite{nbv}. The dissociation barrier, if present, is
basically the DSD migration barrier, which indicates that these
complexes should dissociate very easily at moderate temperatures. This
was confirmed by a simulation performed at 300~K, with a supercell
containing a pair of RR complexes in each dislocation, separated by a
distance of 34.6\AA. On the time scale of 1 ps, one of the kink
complexes undergoes the DSD-emission reaction RR $\rightarrow$ RL +
DSD, with the DSD propagating rather easily towards the other RR
complex, where a DSD + RR $\rightarrow$ LR process takes place.
Overall, a dislocation containing a pair of RR complexes relaxes into
one containing alternating RL and LR kinks, by means of DSD emission
(absorption) and propagation.

\section{Comparison with experimental results}
\label{sec5} 

In Table III we summarize our results for the formation energies and
migration barriers of kinks in the 90$^\circ$ and 30$^\circ$ partial
dislocations. For the 30$^\circ$ partial, of the two equilibrium states
of each kink [(LK,LK$'$) and (RK,RK$'$)], one is to be regarded as an
intermediate metastable state in the propagation of the kink, given
the substantial difference in formation energy between the two states.
Only the state with the lower formation energy will determine
the kink concentration in each case (this lower formation energy is
the number included in Table III). For comparison, results from
Ref.~\onlinecite{bulatov} are also included, as are the ranges
of experimental results for both
quantities, obtained from different
techniques.\cite{gotts,farber,nikit,kolar} We observe that, for the
30$^\circ$ partial, our values are in excellent agreement with the
experimental ones.

\begin{table}
\caption{
Formation energy and migration barriers of dislocation kinks
in Si, in eV. Range of available experimental estimates of is
included. For comparison, results from
Ref.~\protect\onlinecite{bulatov} are also include.
}
\begin{tabular}{lccc}
Dislocation   &Kink type &Formation energy &Migration barrier\\
\hline
30$^\circ$     &LK  &0.35 (0.82\tablenote{From
Ref.~\protect\onlinecite{bulatov}})  &1.53 (0.82\tablenotemark[1]) \\
30$^\circ$     &RK  &1.24 (0.82\tablenotemark[1])
                                     &2.10 (0.74\tablenotemark[1]) \\
90$^\circ$     &LR  &0.12  &1.62 \\
90$^\circ$     &RL  &0.12  &1.62 \\
\hline
Experiments    &    &0.4-0.7 &1.2-1.8
\end{tabular}
\end{table}

The interpretation of these experiments is done according to the
theory of Hirth and Lothe. In this theory, the dislocation velocity is
given by
\begin{equation}
v_d \propto 2 \times \exp \left[-{1 \over kT} \left( U_k 
+ W_m\right)
  \right]\;,
\end{equation}
where $U_k$ is the kink formation energy and $W_m$ is the kink migration
barrier.  This equation is written under the assumption that the two
kinks that result from the nucleation of a stable double kink (a
kink-antikink pair) are equivalent. This assumption does not hold
for the 30$^\circ$ partial, where the left and right kinks are
intrinsically different. The more general form
\begin{eqnarray}
v_d &\propto& \exp \left[-{1 \over 2kT} \left(U_{LK} +  U_{R
K} \right) \right] \nonumber \\
&\times& \left[ \exp \left( -{W^{LK}_m\over kT} \right) + \exp 
\left( -{W^{RK}_m \over kT} \right) \right]\;,
\end{eqnarray}
must be used. We note that the quantity of interest in the first
activated term is the average formation energy of the two kink
species. The second term is derived from the kink velocities, and
therefore the relative velocity appears in the generalized form.  In
the 30$^\circ$ partial this term is dominated by the velocity of the
left kinks (fast carriers), given the much higher migration barrier of
the right kinks (slow carriers). We should point out that the average
formation energy of the kink-antikink pairs in Table III falls within
the range of the experimental numbers, for the 30$^\circ$ partial. As we
mentioned in the introduction, another theory of dislocation glide has
been proposed,\cite{obst1,obst2} in which the motion is controlled by
the pinning of kinks by strong obstacles along the dislocation line,
and the kink migration barriers are not rate controlling. Despite the
fact that our work does not address such pinning mechanisms, and thus
cannot clearly decide between these two theories, our results are
certainly consistent with the HL interpretation.

Strictly speaking, our comparison is only valid for the 30$^\circ$
partial, since we did not consider the true ground state for the
90$^\circ$ partial. In the latter case, the excellent agreement we
obtain for the kink barriers appears to be fortuitous.
Nevertheless, our results are qualitatively consistent with the
experimental images in Ref.~\onlinecite{kolar}, which show a higher
concentration of kinks in the 90$^\circ$ partial. In Table III, we see
that kink energies are lower in this dislocation, as compared to the
30$^\circ$ partial. Obviously, this is only plausible to the extent
that this general trend of lower kink energies carries over to the
ground-state DP structure of the 90$^\circ$ partial.

\section{Conclusions}
\label{sec6} 

In this work, an extensive study of the core reconstruction and
structural excitations in the cores of both the 30$^\circ$ and the
90$^\circ$ partial (in its SP reconstruction) in Si, was presented.
For both partials, we find the core to undergo strong bond
reconstruction, restoring the fourfold coordination of the core
atoms. The reconstructed bonds are stretched by only $\sim$3\% with
respect to bulk values, and the core energies are mostly associated
with the bond angle distortions present in the reconstructed cores.

In the case of the SP structure of the 90$^\circ$ partial, the RD (or
DSD) is associated with a switch of direction of the reconstructed
bonds, and is found to be highly mobile. Kink-DSD complexes are found
to be only marginally stable against emission of a DSD, a reaction
that is observed to proceed rather quickly in our simulations at room
temperature. The LR and RL kinks have very low formation energy,
indicating that they introduce little additional on the SP core, a
result which is consistent with the lower energy of the DP core, as
proposed in Ref.~\onlinecite{bnv}.

For the 30$^\circ$ partial, two kink species (RK and LK) are
identified, and the RK's are found to have higher formation energies
that the LK ones. This is explained by the medium-range behavior of
the associated strains. The RD (or PSD) is related to a
phase-switching of the core reconstruction, and binds strongly with
kinks to form PSD-kink complexes.  These are the more likely sites for
unpaired electrons in the 30$^\circ$ partial core.  The results for
this particular dislocation can be directly compared with experiment,
and we find good agreement between our calculated values for the kink
formation energies and migration barriers and the experimental
results.

\acknowledgments 
Partial support was provided by the DoD Software
Initiative.  J.B. and D.V. acknowledge support from NSF Grants
DMR-91-15342 and DMR-96-13648.


\begin{references}
\bibitem{foot-sol} Reconstruction defects in the core of the
30$^\circ$ and 90$^\circ$ partials have commonly been termed as
solitons, but variants such as antiphase defect have also been used. In
this paper, we shall refer to these as phase-switching defect, in
the case of the 30$^\circ$, and direction-switching defect in the case
of the 90$^\circ$ partial, as explained in the text.
\bibitem{foot-path} The reaction path is the path joining the two
minima in configuration space, along which the gradient is always
parallel to the path. Physically, it is the minimal-action path that
the system would follow to reach a minimum once released from the
saddle-point configuration.
\end{references}
\end{document}